# Learning Priors in High-frequency Domain for Inverse Imaging Reconstruction

Zhuonan He#, Jinjie Zhou#, Dong Liang, *Senior Member, IEEE*, Yuhao Wang, *Senior Member, IEEE*, Qiegen Liu, *Member, IEEE*

*Abstract*—Ill-posed inverse problems in imaging remain an active research topic in several decades, with new approaches constantly emerging. Recognizing that the popular dictionary learning and convolutional sparse coding are both essentially modeling the high-frequency component of an image, which convey most of the semantic information such as texture details, in this work we propose a novel multi-profile high-frequency transform-guided denoising autoencoder as prior (HF-DAEP). To achieve this goal, we first extract a set of multi-profile high-frequency components via a specific transformation and add the artificial Gaussian noise to these high-frequency components as training samples. Then, as the high-frequency prior information is learned, we incorporate it into classical iterative reconstruction process by proximal gradient descent technique. Preliminary results on highly under-sampled magnetic resonance imaging and sparse-view computed tomography reconstruction demonstrate that the proposed method can efficiently reconstruct feature details and present advantages over state-of-the-arts.

*Index Terms*—Imaging reconstruction, high-frequency component, denoising autoencoder network, proximal gradient descent, MRI reconstruction, sparse-view CT.

## I. INTRODUCTION

Inverse imaging reconstruction aims to yield high-quality images from incomplete observations. There are many important applications in medical imaging [1-10], such as magnetic resonance imaging (MRI) and X-ray computed tomography (CT). In MRI, short acquisitions lead to severe degradations of image quality, while long acquisitions may cause motion artifacts, patient discomfort, or even patient harm, therefore, it is crucial to seek a balance between acquisition time and image quality. In X-ray CT, due to radiation exposure, it may pose a potential risk of cancer or genetic disease, reducing the amount of X-ray doses is necessary. Inverse imaging reconstruction with regularization provides a way to mitigate these problems. Its desired solution can be generally modeled as:

$$u = \arg\min_{u} F(Hu, y) + \lambda R(u) \quad (1)$$

This work was supported in part by National Natural Science Foundation of China under 61871206, 61661031 and project of innovative special funds for graduate students (CX2019075, YC2019-S052). # Z. He and J. Zhou contribute equally.

Z. He, J. Zhou, Y. Wang and Q. Liu are with the Department of Electronic Information Engineering, Nanchang University, Nanchang 330031, China. ({hezhuonan, zhoujinjie}@email.ncu.edu.cn, {wangyuhao, liuqiegen}@ncu.edu.cn).

D. Liang is with Paul C. Lauterbur Research Center for Biomedical Imaging, Shenzhen Institutes of Advanced Technology, Chinese Academy of Sciences, Shenzhen 518055, China (dong.liang@siat.ac.cn).

where $H$ represents some linear direct operators, such as a partially observed k-space measure (Fourier transform) or tomographic projection (Radon transform). Regularization term $R(u)$ is adopted to constrain the solution space and plays a critical role in searching for high-quality solutions [11-12]. The desirable solution aims to minimize a linear combination of data-fidelity term $F(Hu, y)$ and regularization term weighted by parameter $\lambda$.

There are many classic regularization models, such as Tikhonov regularization [13], total variation (TV) [14-15], sparse representation [16-17] and nonlocal self-similarity inspired regularizers [18-19]. Both Tikhonov regularization and TV are based on the assumption that image is locally smooth except edges. More specifically, they are good at characterizing the piecewise constant signals and utilizing local structural patterns. Sparse representation algorithms are more effective in representing local image structures, using a few elemental structures from off-the-shelf transforms or learned dictionaries [20-22]. Particularly, dictionary learning (DL) methods select atoms from dictionary and convolutional sparse coding (CSC) methods use convolution operator to sparsely model the images from learned convolutional filter sets. Motivated by the observation that images contain rich repetitive structures, nonlocal self-similarity regularization exploits relations between different patches to facilitate the reconstruction process. Abovementioned methods have excellent reconstruction performance and reasonable computational complexity.

Recently, deep convolution neural networks (CNNs) have shown a tremendous success in solving imaging reconstruction problems [23-25]. However, they lack flexibilities in different tasks. To overcome this weakness, many algorithms have been developed [26-30]. Zhang *et al.* [26] and Bigdeli *et al.* [27] trained networks from noisy-full sampled data pair and then employed them to general image restoration tasks. Similarly, the strategy was also adopted in the work of Tezcan *et al.*, where they trained a variational autoencoder (VAE) network on patches of fully sampled MR images and used the captured distribution as prior for image reconstruction [28].

Built on the observation that both DL and CSC extract the high-frequency component from image, model it and then turn the estimation back to image, we reveal that the image decomposition and estimation summation step can be viewed as a special approximation of the conventional forward transformation and inverse transformation steps. Furthermore, we propose a multi-profile high-frequency strategy as an efficient fashion to approximate the transformation. By integrating it into the recent denoising autoencoder (DAE) prior [27, 29, 30], we present a novel multi-profile high-frequency transform-guided denoising autoencoder as prior (HF-DAEP) to solve imaging problems.

The contributions of this work are summarized as follows:

- By analyzing the representation process of DL and CSC, we reveal that the image decomposition and summation can be approximately represented as forward transformation and inverse transformation, i.e., the image decomposition is defined as a forward transformation, while the inverse transformation is approximately obtained by image estimation summation.
- Multi-profile high-frequency information obtained by decomposition under different scales is utilized as input of DAE network. After the network is trained, the associated learned prior is employed into the conventional iterative reconstruction process. It is demonstrated that HF-DAEP in MRI and CT imaging situations is comparable to or performs better than the state-of-the-arts.

The remainder of the paper is organized as follows. Section II provides a brief description of preliminary work with regard to prior regularization and DAE prior. Section III presents HF-DAEP model and its corresponding iterative solver. Extensive experimental comparisons among HF-DAEP and state-of-the-art methods are conducted in Section IV. Finally, discussions, concluding remarks and directions for future research are given in Section V and VI, respectively.

## II. PRELIMINARY

### A. Prior Regularization

Due to the ill-posed nature of imaging reconstruction, regularization-based techniques have been widely used by constraining the solution space. Thus, it is of great importance to find and model the appropriate prior knowledge [14], [31-41]. For instance, Ying et al. [31] used Tikhonov regularization for improving the reconstruction quality in parallel imaging. Meanwhile, they systematically analyzed the impact of regularization parameters. Similarly, in [14], TV was introduced by Rudin-Osher and Fatemi (ROF) as a regularization approach. The model has been proven powerful in handling edges of degraded image, while it tends to smooth image details and fine structures, resulting in staircase artifacts and contrast losses [32-33].

As an alternative, sparse representation has led to impressive results for various imaging reconstruction applications [34-38]. Ravishankar and Bresler exploited adaptive patch-based dictionary to reconstruct MR images from highly undersampled k-space data [34]. In [36], the authors used K-SVD algorithm [20] to learn an overcomplete dictionary and successfully applied in low-dose CT imaging. Liu et al. [21] proposed a gradient-based DL method for CT imaging, which overcame the drawback of cartoon-like effects in TV regularization. Since most DL methods are patch-based and the learned features often contain shifted versions of the same features, Zeiler et al. proposed CSC representation to alleviate these deficiencies [16]. Recently, Bao et al. used a variant of CSC regularization to cope with sparse-view CT reconstruction [41].

The rapid development of CNNs suggests huge potential in medical imaging. They could be roughly categorized into two types [28-30, 39]. The first category consists of supervised learning approaches, which learn different mappings from training data under specific tasks. Specially, the prior information is learnt implicitly from data, without having to specify them in training objective. For example, Chen et al. [39] proposed a residual encoder-decoder CNN (RED-CNN) for low-dose CT reconstruction. Another category is composed of unsupervised learning approaches. During network training, it designs to learn the probability distribution of the images to be reconstructed. After that, the network-learned image priors are acted as an explicit constraint in various imaging reconstruction tasks [28-30].

### B. Denoising Autoencoder Prior (DAEP)

Autoencoders are typically used for unsupervised representation learning [42]. Especially, Alain and Bengio [43] found that the output of an optimal DAE is a local mean of the true data density. Moreover, Bigdeli and Zwicker [27] used the magnitude of the autoencoder error as a prior. Specifically, by assuming DAE to be $A$ and its output is $A_{\sigma_\eta}(u) = A(u+\eta)$, then DAE can be trained by minimizing the loss function:

$$L_{DAE} = E_{\eta,u}[\|u - A_{\sigma_\eta}(u)\|^2] \quad (2)$$

where the expectation $E_{\eta,u}[\circ]$ is conducted overall images $u$ and Gaussian noise $\eta$ with standard variance $\sigma_\eta$. Aditionally, $A_{\sigma_\eta}(u)$ is related to the true data density $p(u)$ as follows:

$$A_{\sigma_\eta}(u) = \frac{\int (u-\eta) g_{\sigma_\eta}(\eta) p(u-\eta) d\eta}{\int g_{\sigma_\eta}(\eta) p(u-\eta) d\eta} \quad (3)$$

DAEP utilizes the migratory characteristic of prior information and uses the magnitude of this mean shift vector as the negative log-likelihood of image prior, i.e.,

$$\|A_{\sigma_\eta}(u) - u\|^2 = \|\sigma_\eta^2 \nabla \log[g_{\sigma_\eta} * q](u)\|^2 \quad (4)$$

Furthermore, Li et al. [29] adopted multi-channel and multi-model strategy to enhance the naïve DAEP for grayscale image restoration tasks, dubbed as MEDAEP. It prefers to avoid getting stuck in local minima and makes the iterative process to be more robust. In particular, they used color images $\{U | U = [u_r, u_g, u_b]\}$ as training samples to train DAE $A_{\sigma_\eta}$. Accordingly, the autoencoder error in MEDAEP is

$$\|A_{\sigma_\eta}(U) - U\|^2 = \|\sigma_\eta^2 \nabla \log[g_{\sigma_\eta} * q](U)\|^2 \quad (5)$$

As can be seen in Eqs. (4) and (5), the effectiveness of prior (i.e., the autoencoder error) depends on the distribution of training data, and is proportional to the gradient of its log-density. In this work, we exploit a novel transform on the basis of DAE to form a more efficient prior in high-dimensional and high-frequency domain.

## III. PROPOSED METHOD

As well known, DL and CSC are two popular sparse representation methods. Besides of the core representation step, there exist a pre-process and a post-process step: high-frequency component extraction and estimation component summation. In this section, after reviewing the whole procedure implemented in DL and CSC, we reveal an important phenomenon, i.e., if decomposing the image into low-frequency and high-frequency components is defined as a forward transformation, then the associated components summation procedure can be interpreted as an approximation of a backward transformation. This observation paves a new direction of how to exploit prior in a better way. Therefore, by means of DAEP, a novel multi-profile high-frequency strategy acting as a better approximation to the transformation, coined HF-DAEP, is derived.

## A. Motivation

Owing to the fact that high-frequency component can effectively represent the image textures and details, extracting high-frequency component is used as an important operator for exploiting prior. In DL and CSC methods, many researchers have integrated them into reconstruction tasks and achieved outstanding performance [44-48].

DL aims to reduce dimensionality subspace by finding an appropriate dictionary for data representation. The dictionary $D$ trained on image patches helps to exploit the sparsity prior of images. Mathematically, given a matrix $X = [\cdots, \vec{u}_i - \bar{u}_i, \vec{u}_{i+1} - \bar{u}_{i+1}, \cdots]$ that each column corresponding to one patch, DL trains a dictionary via:

$$\min_D R(D | u) \equiv \|X - DY\|_F^2 + \hat{\lambda}\|Y\|_1 \quad (6)$$

where $\hat{\lambda}$ is a standard Lagrangian multiplier, $Y$ stands for sparse coding matrix of all patches. $u_i$ is the $i$-th image patch vector with size $\sqrt{n} \times \sqrt{n}$ and $\bar{u}_i$ is the mean of $\vec{u}_i$. Three steps in DL can be summarized as follows: (i) A matrix $X$ is formed by arranging each training patch as a column, where mean-subtract is conducted; (ii) The dictionary $D$ is learned by pursuing solution from Eq. (6); (iii) $\hat{X} = [\cdots, D\vec{y}_i + \bar{u}_i, D\vec{y}_{i+1} - \bar{u}_{i+1}, \cdots]$ is returned by summing the matrix factorization estimation and the means of patches. Finally, followed by reassembling image, region-averaging between overlapped patches is done to obtain the final estimated image. Visual illustration of the above process is shown in Fig. 1.

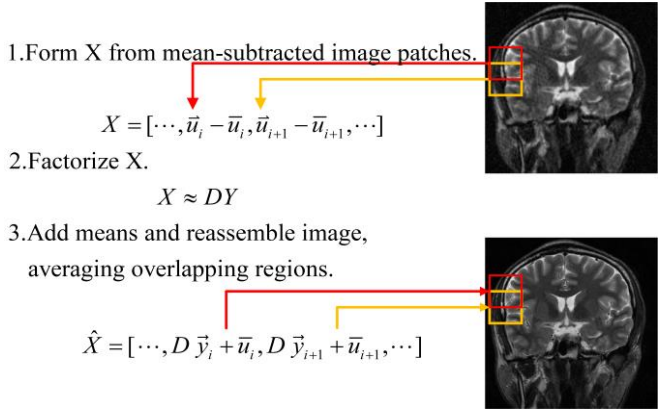

Fig. 1. Visual illustration of obtaining high-frequency image components for dictionary learning.

In addition, Zeiler et al. [16] proposed a convolutional implementation of sparse coding to encode the whole image, which effectively utilizes the consistency of overlapping patches. Traditionally, CSC is expressive of the following minimization:

$$\min_d R(d|u) \equiv \left\|u - \sum_{k=1}^{K} d_k * z_k\right\|_2^2 + \beta \sum_{k=1}^{K}\|z_k\|_1 \quad (7)$$
$$s.t. \ \|d_k\|_2^2 \leq 1, \ k = 1, \cdots, K$$

where $\{z_k\}$ are sparse maps that approximate $u$ when convolved with the corresponding filters $\{d_k\}$ of fixed spatial support. Here, $*$ is 2D convolution operator and $\beta$ is a positive regularization parameter.

Similar to DL, CSC can be summarized as follows: First, there is a low-frequency feature map $u_l$ generated from input image $u$ by a low-pass filter. Second, the remaining high-frequency component $u_h = u - u_l$ is sparsely modeled by the learned filters. Last but not least, the recovery image is obtained by summing the low-frequency component and the estimated high-frequency component. This whole process is illustrated in Fig. 2.

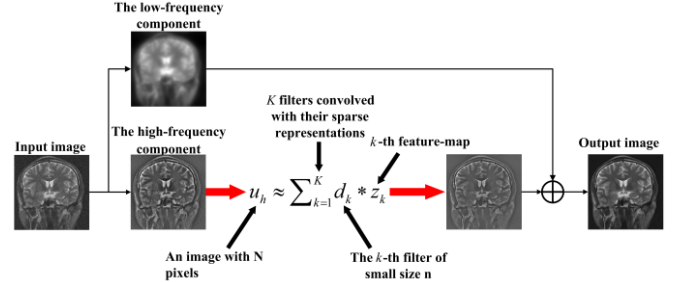

Fig. 2. Visual illustration of extracting high-frequency image component for filters learning in convolutional sparse coding.

All in all, after decomposing image into low-frequency and high-frequency parts, both DL and CSC exploit prior in high-frequency domain. They obtain the final reconstruction results by summing low-frequency part and the new estimated high-frequency part at the third step. This phenomenon indicates that utilizing high-frequency information indeed facilitates to image reconstruction.

## B. Interpretation of Forward and Inverse Transform

In fact, image decomposition and summation in DL and CSC can be seen as approximations of forward and inverse transform. In the following context, we reveal that both DL and CSC are special implementations of the transformations.

Specifically, each image patch in DL has been transferred to an object in high-frequency domain by removing the mean components, i.e.,

$$u_{h,i} = u_i - Mean(u_i) \quad (8)$$

By means of convolution operator, Eq. (8) can be rewritten as follows:

$$u_l = M * u; \quad u_h = u - u_l \quad (9)$$

where $M$ is a matrix whose elements are all ones. Clearly, $M * u$ is equal to applying a low-frequency filter to the image.

Additionally, the low-frequency component $u_l$ in CSC is first computed via Tikhonov regularization and then the high-frequency component is obtained via $u_h = u - u_l$, i.e.,

$$u_l = \arg\min_{u_l} \|u_l - u\|_2^2 + \alpha\|\nabla u_l\|_2^2 = (I + \alpha\nabla^T\nabla)^{-1}u$$
$$u_h = u - u_l \quad (10)$$

where $\nabla u_l$ denotes the image gradient, $\alpha > 0$ regulates the amount of high-frequency component in $u_l$.

As can be seen, Eqs. (9) and (10) are two special formulations of the low-pass filtering operator in $u_l = Lu$, i.e., $L = M$ and $L = (I + \alpha\nabla^T\nabla)^{-1}$). In summary, we can define a forward operator $W$ that transfers the image from the whole image domain to partly high-frequency domain for prior modeling, i.e., $u_h = u - u_l = (1 - L)u = Wu$.

Accordingly, we reveal that the component summation operator $u = u_h + Lu$ in DL and CSC approximately acts as a backward transformation $W^{-1}$ that returning the estimation in high-frequency domain back to image domain, i.e.,

$$u = u_h + Lu \approx u_h + Lu_h$$
$$= (I+L)u_h \approx (I-L)^{-1}u_h \equiv W^{-1}u_h \quad (11)$$

Two conclusions can be achieved as follows:

First, Eq. (11) is obtained by two approximations, i.e., $Lu_h = L(u-u_l) = L(u-Lu) \approx Lu$ and Taylor series expansion approximation $(I-L)^{-1} \approx I+L+o(L^2)$. They become to be equal only if $L \to 0$. Specifically, when $L \to 0$, it yields $u_h \to u$ and $W \to I$, subsequently Eq. (11) tends to be more accurate.

Second, unlike the conventional transformation (e.g., wavelet transform) that has exact inverse transformation, the inverse process in Eq. (11) is obtained by approximated transformation representation. Therefore, this is an essential drawback existed in DL and CSC even if the representation modeling of high-frequency component is satisfied. On the other hand, since low-frequency and high-frequency components belong to different orders of magnitude, the high-frequency component extraction occured in DL and CSC in favor of subsequent representation modeling.

As analyzed above, how to balance the representation capability of high-frequency component and accuracy of inverse transformation is urgent task. In this work, we propose a new strategy to execute the forward and inverse transformation. Specifically, we extend them via conducting a series of low-frequency operators $\{L_i\}$. Furthermore, we will show that employing the tool DAEP on the resulting high-frequency components are also very effective.

### C. Proposed HF-DAEP

In the proposed HF-DAEP, the forward transform $W$ consists of the following manipulations: First, we set different regularization parameters $\{\alpha_i\}$ in low-pass filters $\{L_i\}$; Second, after subtracting the high-frequency components, a series of high-frequency components on different profiles is obtained and the elements at the same spatial location are stacked as a tensor; Finally, the resulting tensor is used as the network input. Concretely, the sets of low-frequency components $\{u_{l,\alpha_i}\}$ and high-frequency component $\{u_{h,\alpha_i}\}$ are obtained similarly as in Eq. (10), i.e.,

$$u_{l,\alpha_i} = (I+\alpha_i \nabla^T \nabla)^{-1}u; \quad u_{h,\alpha_i} = u - u_{l,\alpha_i} \quad (12)$$

By setting $\alpha_i$-value to be very large, $u_{h,\alpha_i}$ approaches to the input image $u$; Meanwhile, by setting $\alpha_i$-value to be relatively small, then $u_{h,\alpha_i}$ is very similar to that in CSC. Since high-frequency components in Eq. (12) are generated with different amounts, they have unique respective characteristics. It is vital to integrate them for expoiting multi-profile representation and priori information, such as as to benefit from image domain and high-frequancy domain simultaneously, thus promoting reconstruction performance.

To better understand the visual characteristics among different high-frequency profiles, a demonstration of high-frequency components $W(u) = Concat[\{u_{h,\alpha_i}\}]$ from $\{\alpha_i | i=1,\cdots,5\}$-values is visualized in Fig. 3, where $Concat[\circ]$ is a function for channel-wise concatenation operation. Particularly, in the case of $\alpha$-value to be infinity, the resulting high-frequency component is equal to the original image. As $\alpha$-value decreases, the edges and details of image will be more prominent, while the global information is less contained. More details for illustrating the forward transform $W$ is shown in Fig. 4(a).

We employ the RED-Net [49] architecture as the DAE network $A_{\sigma_\eta}$ for exploiting multi-profile high-frequency prior. Inspired by the work in [29], the HF-DAEP prior is denoted as:

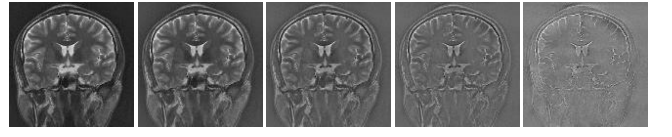

Fig. 3. Visual illustration of high-frequency components. From left to right: Decomposed image obtained by setting lambda 5000, 500, 50, 12 and 1.

$$R(u) \equiv \left\| W(u) - A_{\sigma_\eta}(W(u)) \right\|^2 \quad (13)$$

where $W$ stands for the transform operator to obtain multi-profile high-frequency features. Similar to Eq. (2), the artificial Gaussian noise $\eta$ is added to $W(u)$ as the input of network.

We inherit the spirit in DL and CSC that using image decomposition and component summation to approximate the forward and inverse transformations. Specifically, there exists two important steps in implementing Eq. (13): Forward operator $W(u) = Concat[\{u_{h,\alpha_i} | i=1,\cdots,N\}]$ shown in Fig. 4(a) and the associated backward operator $W^{-1}[\{u_{h,\alpha_i}\}] = Mean[\{u_{h,\alpha_i}+u_{l,\alpha_i}\}] = u$ depicted in Fig. 4(b). $N$ stands for the number of filters. Specifically in the approximated backward operator, the updated high-frequency estimation adds with the latest low-frequency components, followed by mean operation to get the final result.

### D. Solver of HF-DAEP

The general mathematical HF-DAEP model for imaging reconstruction is as follows:

$$\min_u F(Hu,y) + \lambda \left\| W(u) - A_{\sigma_\eta}(W(u)) \right\|^2 \quad (14)$$

As discussed in the previous subsections, the second term in Eq. (14) tackles with network-driven prior information. We employ proximal gradient descent method [50] to tackle the nonlinear model. Subsequently, Eq. (14) is approximated as follows:

$$\min_u F(Hu,y) + \frac{\lambda}{\beta} \left\| u - (u^k - \beta \nabla G(u^k)) \right\|^2 \quad (15)$$

where $\nabla G(u) = W^{-1}\{[1-\nabla A_{\sigma_\eta}^T(W(u))][W(u)-A_{\sigma_\eta}(W(u))]\}$ and $G(u) = \left\| W(u) - A_{\sigma_\eta}(W(u)) \right\|^2$. Notice that the parameters in $A_{\sigma_\eta}(\circ)$ have already been learned at network training stage. $\nabla A_{\sigma_\eta}^T(\circ)$ is the derivative which can be solved by calculating backpropagation of the network. Here, we set $\beta=1$ through empirical observation and it works well in all experiments. Especially, the data fidelity term $F(Hu,y)$ has corresponding representations for specific tasks, such as MRI and CT reconstruction.

**MRI reconstruction:** In MRI, the partially observed measurement can be written as $y=Hu+n$, where $H$ represents the partially Fourier encoding matrix and $n$ is the noise in Frequency domain. By considering the linear con-

straint in data consistency, Eq. (15) can be modeled as:

$$\min_u \|Hu - y\|^2 + \lambda \|u - (u^k - \nabla G(u^k))\|^2 \quad (16)$$

Then, Eq. (16) can be solved by setting its gradient to be zero, i.e.:

$$u^{k+1} = \frac{H^T y + \lambda[u^k - \nabla G(u^k)]}{(H^T H + \lambda)} \quad (17)$$

We update the solution $u^{k+1}$ by alternatively calculating prior gradient and least-square (LS) solver until its value converges.

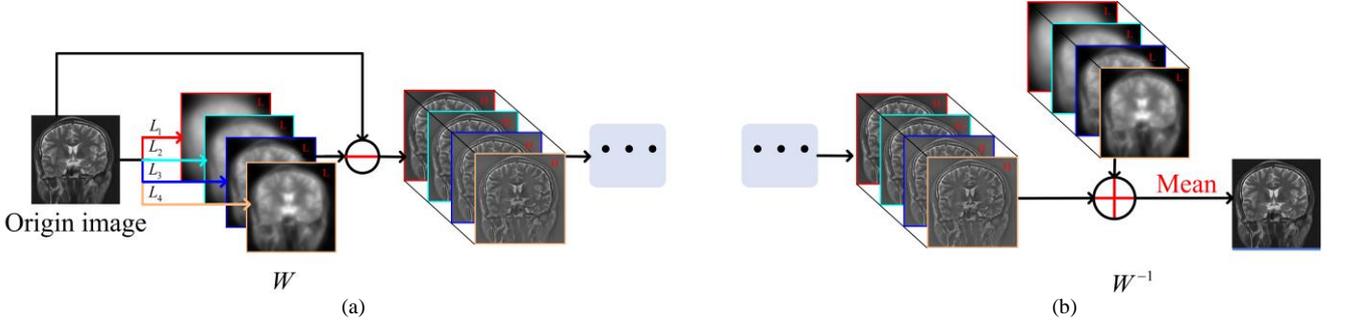

Fig. 4. Demonstration of (a) the forward transform operator $W$ and (b) the backward transform operator $W^{-1}$ in HF-DAEP.

**CT reconstruction**: Due to the attenuation property, the calibrated and log-transformed projection data follow approximately a Gaussian distribution. It has an associated relationship between the data sample mean and variance, i.e., [51-52]:

$$\delta_i^2 = f_i \exp(\mu_i / T) \quad (18)$$

where $\mu_i$ is the mean and $\delta_i^2$ is the variance of projection measurements at bin $i$. $T$ is a scaling parameter and $f_i$ is a parameter adaptive to different detector bins. Based on the above basic properties, the penalized weighted least-squares (PWLS) based sparse-view projection reconstruction can be modeled as follows [53]:

$$\min_u (y - Hu)^T \Sigma^{-1} (y - Hu) + \lambda \|u - (u^k - \nabla G(u^k))\|^2 \quad (19)$$

where $H$ denotes the system matrix with a size of $I \times J$ ($I$ is the total number of projection data and $J$ is the total number of image pixels). $\Sigma$ is a diagonal matrix with the $i$-th element of $\delta_i^2$. With the separable paraboloid surrogate method [54], the solution of Eq. (19) can be achieved by:

$$u_j^{k+1} = u_j^k - \frac{\sum_{i=1}^{I}((h_{ij}/\delta_i)([Hu^k]_i - y_i))}{\sum_{i=1}^{I}((h_{ij}/\delta_i)\sum_{z=1}^{J} h_{iz}) + \lambda} - \frac{\lambda \nabla G(u_j^k)}{\sum_{i=1}^{I}((h_{ij}/\delta_i)\sum_{z=1}^{J} h_{iz}) + \lambda}, \quad (20)$$
$$j = 1, 2, \cdots, J$$

where $k = 1, 2, \cdots, K$ and $h_{ij}$ stand for the iteration index and the element of system matrix $H$, respectively.

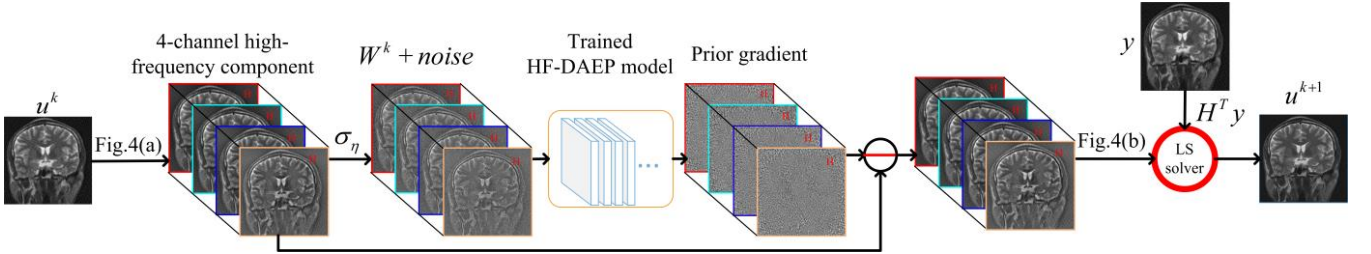

Fig. 5. Illustration of HF-DAEP at iterative reconstruction phase. Here MRI reconstruction is visualized.

In summary, the mathematical models in both MRI and CT reconstruction are tackled by the proximal gradient descent and alternating optimization. Fig. 5 visualizes a schematic flowchart of HF-DAEP for MRI reconstruction. Overall training and testing phase of HF-DAEP are as follows:

| Algorithm 1 HF-DAEP |
|---|
| **Training stage** |
| **High-frequency dataset**: $W(u) = Concat[\{u_{h,\alpha_i} \| i = 1, \cdots, N\}]$ |
| **Outputs**: Trained network $A_{\sigma_\eta}(\circ)$ |
| **Testing stage** |
| **Initialization:** $u^0 = H^T y$ |
| **For** $k = 1, 2, \cdots, K$ |
|     Update variable: $W(u^k) = Concat[\{u^k_{h,\alpha_i}\}]$ |
|     Calculate the components in prior gradient $\nabla G(u^k)$ |
|   **If** MRI reconstruction |
|     Update the solution via Eq. (17) |
|   **Elseif** CT reconstruction |
|     Update the solution via Eq. (20) |
|   **End** |
| **End** |

## IV. EXPERIMENTS

In this section, the performance of HF-DAEP is demonstrated in MRI and CT reconstruction. All implementations are coded in Matlab2016. Computations are performed with Intel(R) Core (TM) i7-7700 central processing unit at 3.60GHz, 64G RAM and GeForce Titan XP. For the purpose of reproducible, source code of HF-DAEP is available at https://github.com/yqx7150/HFDAEP.

### A. Training Datasets and Settings

**MRI reconstruction**: The collected training dataset includes

500 complex-value MR images in this work. Particularly, 400 of them are used for network training and 100 subjects for validation. Concretely, the training patch size and batch size are set to be $40\times40$ and 64, respectively. Noise level in the prior learning stage is set to be $\sigma_\eta=25$. Here, we handle complex data by concatenating the real and imaginary parts as channels, i.e., the real and imaginary components are set as the network input simultaneously. $\alpha$-values of 1000, 800, 400 and 50 (i.e., $N=4$) are used to extract multi-profile high-frequency components in both MRI and CT reconstruction. Moreover, the input of HF-DAEP is converted from $C^{m\times n\times 2}$ space into $R^{m\times n\times 8}$ space.

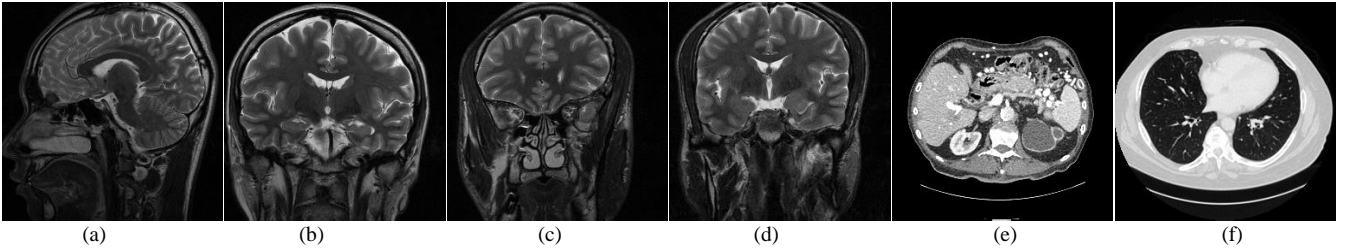

(a) (b) (c) (d) (e) (f)

Fig. 7. Representative testing images. (a)(b)(c)(d) Test 1-4; (e) Abdominal image. (f) Thoracic image. The former four images for MRI reconstruction and the latter two images for CT reconstruction.

**CT reconstruction**: The (DIVerse 2K) DIV2K resolution high quality images [55] are chosen as training dataset. The DIV2K dataset collects 800 high definitions high resolution RGB images with a large diversity of contents, which is originally utilized for the NTIRE challenge on super-resolution at CVPR 2017. In the collected training dataset of our experiment, we exploit 210 images from DIV2K dataset as training set and 50 images for valiation. Specifically, the DIV2K dataset involves samples with more diverse and abundant contexts.

At reconstruction stage, we use a variety of sampling schemes on 31 complex-valued MRI data and two CT images. Fig. 7 displays some fully-sampled representative MR and CT images. For quantitative comparison, peak signal to noise ratio (PSNR), structural similarity (SSIM) and high-frequency error norm (HFEN) are introduced to measure the quality of the reconstruction.

The PSNR is calculated by:
$$PSNR(u,\hat{u}) = 10\log_{10}[\text{Max}(\hat{u})/\|u-\hat{u}\|_2^2] \quad (21)$$
where $\hat{u}$ and $u$ are ground-truth and the reconstructed image, respectively.

The SSIM between $u$ and $\hat{u}$ is defined as:
$$SSIM(u,\hat{u}) = \frac{(2\mu_u\mu_{\hat{u}} + c_1)(2\sigma_{u\hat{u}} + c_2)}{(\mu_u^2 + \mu_{\hat{u}}^2 + c_1)(\sigma_u^2 + \sigma_{\hat{u}}^2 + c_2)} \quad (22)$$
where $\mu_u$ and $\mu_{\hat{u}}$ are mean values of image $u$ and $\hat{u}$. $\sigma_u$ and $\sigma_{\hat{u}}$ denote their standard deviations. $\sigma_{u\hat{u}}$ is covariance of $u$ and $\hat{u}$. Moreover, $c_1 = (K_1C)^2$ and $c_2 = (K_2C)^2$, where $C$ is the dynamic range of pixel intensity. Additionally, $K_1=0.01$ and $K_2=0.03$.

Finally, HFEN quantifies the reconstruction quality of edges and fine features. It is computed as the norm of the result obtained by Laplacian of Gaussian (LOG) filtering:
$$HFEN = \|LoG(u) - LoG(\hat{u})\|_F^2 / \|LoG(\hat{u})\|_F^2 \quad (23)$$
where the filter kernel is of size $15\times15$ pixels and has a standard deviation of 1.5 pixels.

### B. MRI Reconstruction

For compressed sensing MRI (CS-MRI), the k-space undersampled data under acceleration factor $R=5$ and 6.7 are considered. Furthermore, comparison performances under different sampling patterns are also investigated, i.e., the variable density 2D random sampling, pseudo radial sampling and Cartesian sampling. HF-DAEP is compared with several leading CS-MRI methods, i.e., the reference-derived sparse representation method PANO [56], the dictionary learning method DLMRI [34] and FDLCP [47], the low-rank based NLR-CS [57], the end-to-end DC-CNN [58], and the enhanced DAEP (EDAEP) [30].

Table I reports the PSNR, SSIM and HFEN results of the competing methods. It can be observed that HF-DAEP outperforms other methods. For almost undersampling patterns, HF-DAEP works better than EDAEP. Moreover, for 2D random sampling with $R=6.7$, HF-DAEP outperforms EDAEP and NLR-CS by 0.44dB and 2.34dB, respectively.

For visual comparison, Figs. 8-9 depict some reconstruction results by the competing methods. Better visual comparison can be made by zooming the images on the screen. Although the reconstruction images obtained by NLR-CS and DC-CNN achieve much better visual quality than those of DLMRI, PANO and FDLCP, they still suffer from some undesirable artifacts and lost details such as ringing, jaggy and staircase artifacts.

All in all, it can be observed that HF-DAEP produces not only sharp large-scale edges but also fine-scale image details. It preserves the image structures and textures better than other competing methods. Specifically, both EDAEP and HF-DAEP fall into the category of multi-channel derived prior, the superior performance of HF-DAEP may be attributed to the efficient employment of multi-profile high-frequency information.

### C. CT Reconstruction

In this experiment, Siddon's ray driven method [59] was used to generate the projection data in fan-beam geometry. The source-to-rotation center distance was 40cm while the detector-to-rotation center was 40cm. The image region was set to 20cm $\times$ 20cm. The detector width was 41.3cm containing 512 detector elements. There were 360 viewing angles uniformly distributed over a full scan range.

To demonstrate the effectiveness of HF-DAEP in CT image reconstruction, we compare the proposed method with several competitive techniques: sinogram filtering based filtered back-projection method (FBP) [60], CSC-regularized PWLS method (PWLS-CSC) [42], an improved PWLS-CSC version by imposing gradient regularization (PWLS-CSCGR) [42]. To make a fair compari-son among the competing methods, we have carefully tuned their parameters to achieve the best performance.

TABLE I
AVERAGE PSNR, SSIM AND HFEN VALUES OF RECONSTRUCTING 31 TEST IMAGES AT FIVE SAMPLING TRAJECTORIES.

|  | $R=5$, 2D Random | $R=6.7$, 2D Random | $R=5$, Radial | $R=6.7$, Radial | $R=6.7$, 1D Cartesian |
|---|---|---|---|---|---|
| **DLMRI** | 30.47/0.8423/1.16 | 27.63/0.7518/2.02 | 31.21/0.8602/1.10 | 29.36/0.8103/1.58 | 26.50/0.7390/2.51 |
| **PANO** | 32.19/0.8729/0.94 | 29.12/0.7964/1.77 | 32.44/0.8777/0.96 | 30.60/0.8372/1.37 | 27.51/0.7683/2.28 |
| **FDLCP** | 32.75/0.8700/0.77 | 30.14/0.8004/1.44 | 32.97/0.8770/0.80 | 31.31/0.8391/1.13 | 27.91/0.7776/2.15 |
| **NLR-CS** | 33.19/0.8776/0.76 | 30.34/0.8087/1.46 | 33.32/0.8812/0.79 | 31.35/0.8494/1.17 | 28.23/0.7798/2.03 |
| **DC-CNN** | 33.17/0.8790/0.80 | 28.78/0.7873/1.83 | 32.68/0.8791/0.95 | 30.57/0.8348/1.38 | 27.05/0.7506/2.44 |
| **EDAEP** | 33.07/0.8937/0.80 | 30.68/0.8433/1.31 | 33.49/0.8990/0.79 | 32.00/0.8716/1.05 | 28.85/0.8041/1.81 |
| **HF-DAEP** | **33.38/0.8954/0.72** | **31.12/0.8490/1.25** | **33.65/0.8998/0.73** | **32.13/0.8721/1.01** | **28.88/0.8042/1.81** |

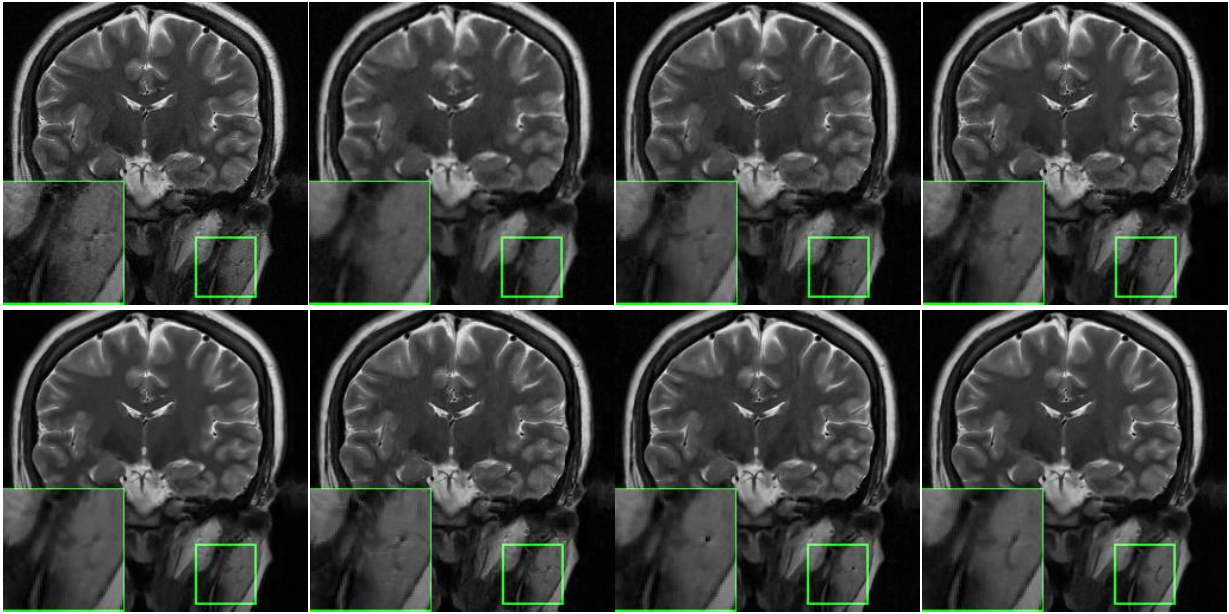

Fig. 8. Visual comparisons under 2D Random sampling at $R=5$. Top line: reference image, reconstruction using DLMRI, PANO and FDLCP; Bottom line: reconstruction using NLR-CS, DC-CNN, EDAEP and HFDAEP.

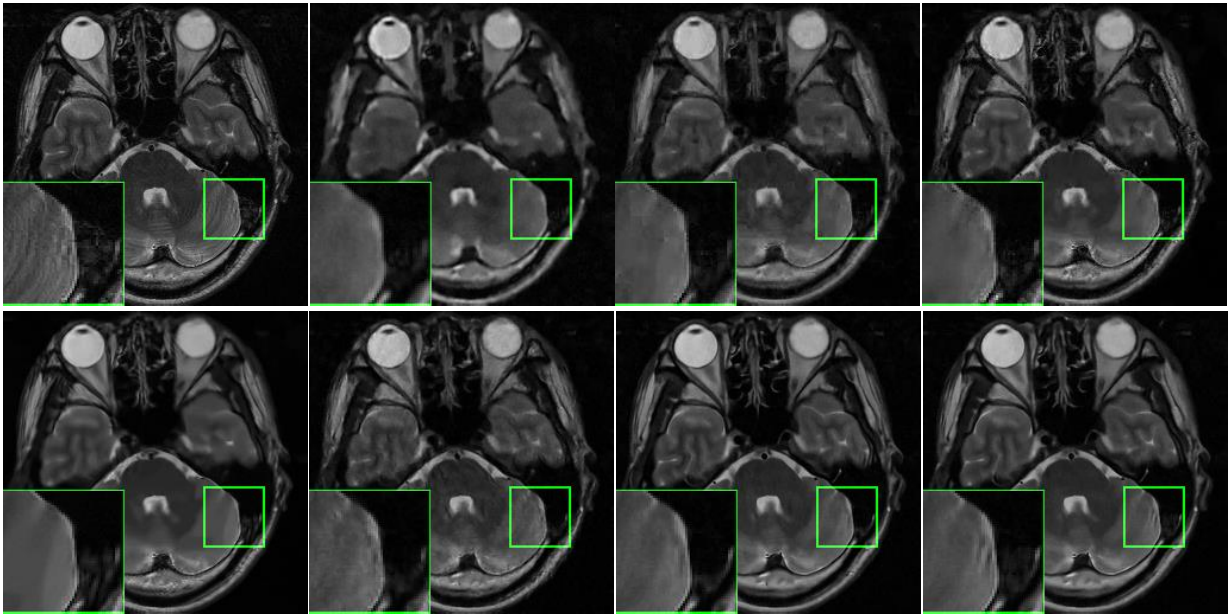

Fig. 9. Visual comparisons under pseudo radial sampling at $R=6.7$. Top line: reference image, reconstructed image using DLMRI, PANO, and FDLCP; Bottom line: reconstructed image using NLR-CS, DC-CNN, EDAEP, and HFDAEP.

Table II tabulates the quantitative evaluations for abdominal and thoracic images that reconstructed from 48, 64, 80 projection views with different methods. HF-DAEP obtains the best results under all conditions. Specifically, the average PSNR values of HF-DAEP in abdominal image are over 0.68dB, 0.96dB and 0.55dB than the second-best method under three projection views. Meanwhile, the average PSNR values gain of HF-DAEP are 1.80dB, 1.80dB, and 1.07dB over the second-best method in thoracic image.

Visually, Figs. 10-11 display the reconstruction results and the difference images of HF-DAEP and the other methods in 48 and 64 projection views. It clearly illustrates that HF-DAEP removes most of streaking artifact patterns and preserves details of underlying images in first row. To further demonstrate the gains of HF-DAEP, difference images relative to the original image are shown in second row. It can be seen that the loss of structural information of HF-DAEP is least than other algorithms, the artifacts are well suppressed and image edge loss is effectively reduced.

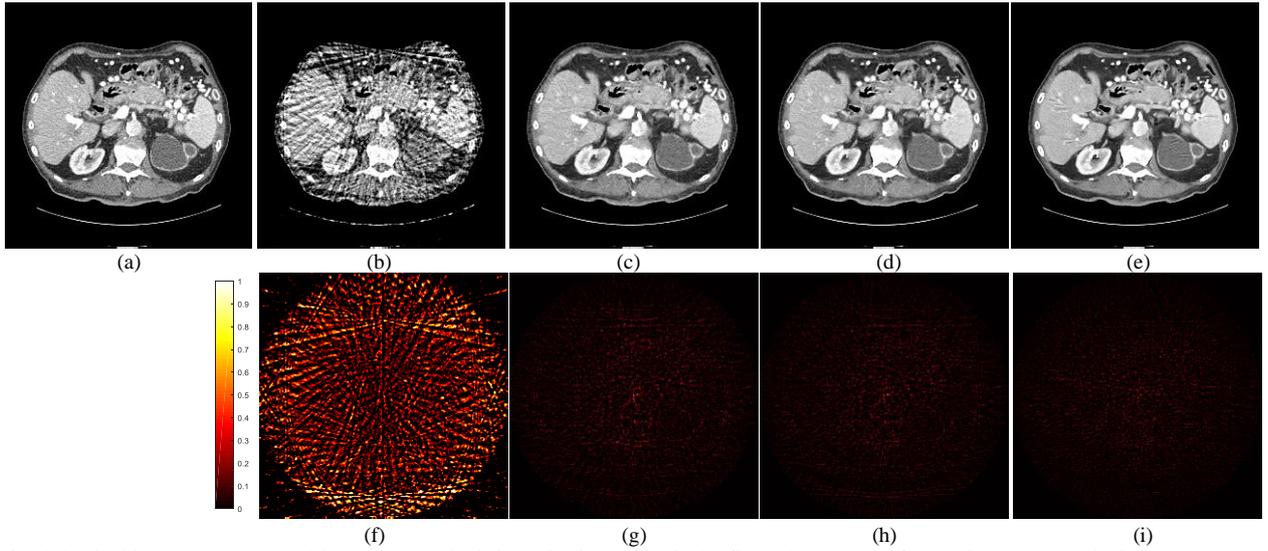

Fig. 10. Abdominal images reconstructed by various methods in projection 64 are in the first column. (a) Reference image versus the images reconstructed by (b) FBP, (c) PWLS-CSC, (d) PWLS-CSCGR and (e) HF-DAEP. The display window is [-150 250]HU. Difference images relative to the original image are in the second column. Results for (f) FBP, (g) PWLS-CSC, (h) PWLS-CSCGR and (i) HF-DAEP. The display window is [-1000 -700]HU.

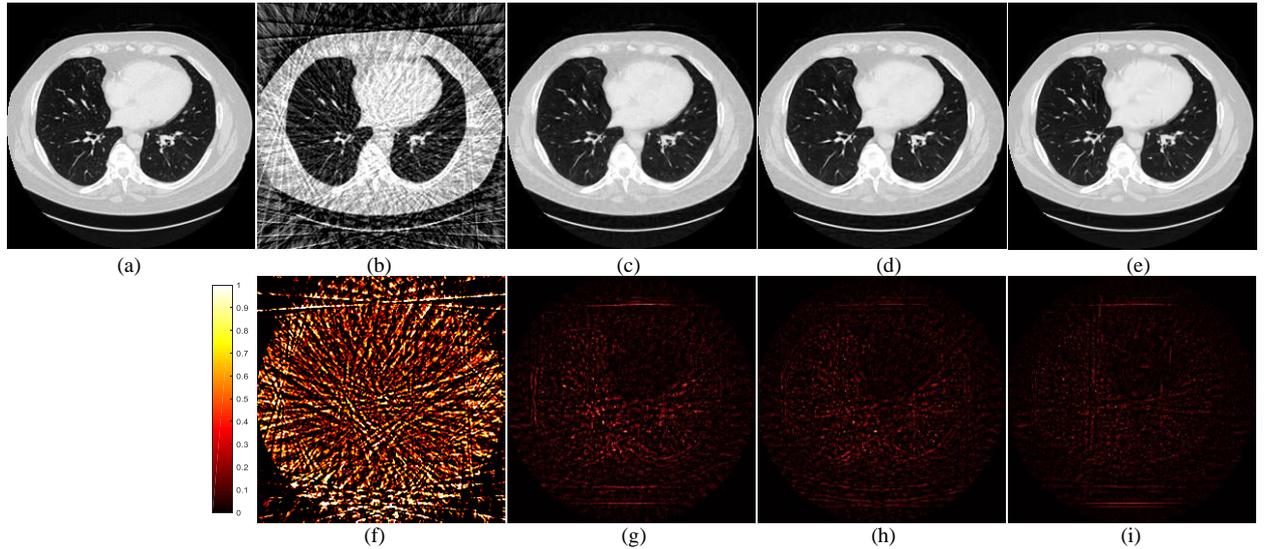

Fig. 11. Thoracic images reconstructed by various methods in projection 48 are in the first column. (a) Reference image versus the images reconstructed by (b) FBP, (c) PWLS-CSC, (d) PWLS-CSCGR and (e) HF-DAEP. The display window is [-1000 250]HU. Difference images relative to the original image are in the second column. Results for (f) FBP, (g) PWLS-CSC, (h) PWLS-CSCGR and (i) HF-DAEP. The display window is [-1000 -700]HU.

TABLE II
AVERAGE PSNR, SSIM VALUES OF RECONSTRUCTING TEST IMAGES AT DIFFERENT PROJECTION VIEWS WITH THE SAME PERCENTAGE.

| Num. of Views | 48 | 64 | 80 |
|---|---|---|---|
| Abdominal | | | |
| FBP | 24.15/0.4138 | 25.35/0.4586 | 26.40/0.5429 |
| PWLS-CSC | 41.36/0.9638 | 44.92/0.9806 | 47.75/0.9891 |
| PWLS-CSCGR | 42.52/0.9692 | 45.73/0.9833 | 48.38/0.9903 |
| HF-DAEP | **43.20/0.9770** | **46.69/0.9867** | **48.93/0.9918** |
| Thoracic | | | |
| FBP | 21.85/0.3623 | 22.99/0.4234 | 23.90/0.4791 |
| PWLS-CSC | 41.55/0.9718 | 45.07/0.9852 | 48.69/0.9927 |
| PWLS-CSCGR | 42.78/0.9769 | 46.07/0.9875 | 49.54/0.9937 |
| HF-DAEP | **44.58/0.9844** | **47.87/0.9915** | **50.61/0.9951** |

## V. DISCUSSIONS

### A. Possible Formulations of Transformation $W$

We consider more possible formulations of $W$, which extracts high-frequency component from image. Similarly, TV provides a way to extract gradient images which focus on the local information. Many researchers have achieved good performances in restoring image details by utilizing TV-based methods [14-15].

$\nabla = [\nabla_x, \nabla_y]$ denotes the gradients of a 2D image $u(j,k)$ in horizontal and vertical directions, $j$ and $k$ are position indices. In discrete domain, it is defined as simple forward difference:

$$\begin{aligned}\nabla_x u(j,k) &= u(j,k+1) - u(j,k) \\ \nabla_y u(j,k) &= u(j+1,k) - u(j,k)\end{aligned} \quad (24)$$

It can be seen that these two gradient component images are obtained by convolution filter with size $1 \times 2$ (or $2 \times 1$). Specially, the difference operators in TV extract high-frequency information to some extent, as shown in Fig. 12. The gradient images in horizontal and vertical directions are similar to high-frequency component obtained by CSC.

Since the difference operator can be seen as a special tool to get high-frequency information, we investigate its effectiveness in constructing $W$ by combining other high-frequency parts, i.e., $W(u) = \{u_{h,\alpha_1}, u_{h,\alpha_2}, u_{h,\alpha_3}, \nabla_x(u)\}$.

Table III lists the results of HF-DAEP and its variant in MRI reconstruction. Particularly, HF-DAEP employs four high-frequency components, and HF-DAEP-II consists of three high-frequency components and a gradient part in Eq. (24). Compared to the reconstruction results of conventional iterative methods in Table I, HF-DAEP-II attains reasonable performance gains. Nevertheless, its performance is inferior to HF-DAEP. This phenomenon indicates that the difference operator is not an ideal tool to extract high-frequency information. In other words, its inverse transform is inaccurate and is consistent with the statement in [21].

TABLE III
AVERAGE PSNR, SSIM AND HFEN VALUES OF RECONSTRUCTING 31 TEST IMAGES

|  | HF-DAEP | HF-DAEP-II |
|---|---|---|
| $R = 6.7$, 2D Random | **31.12/0.8490/1.25** | 30.38/0.8349/1.40 |
| $R = 6.7$, Radial | **32.13/0.8698/1.01** | 31.47/0.8579/1.15 |
| $R = 6.7$, 1D Cartesian | **28.88/0.8011/1.83** | 28.19/0.7895/2.01 |

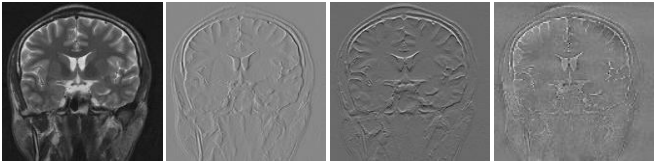

Fig. 12. Visual illustration of gradient images. From left to right: reference image, gradient images in the horizontal and vertical directions, CSC pre-processing image.

### B. Variants of Network

To evaluate the impact of high-frequency combinations $\{\alpha_i\}$, experiments were performed with three different combinations on MRI reconstruction in Table IV. In the circumstances of setting high $\alpha$-value such as {1200, 1000, 800, 400}, the performance of HF-DAEP is close to that of EDAEP. However, setting small $\alpha$-values like {800, 400, 50, 10}, its performance degrades. Therefore, we empirically set $\alpha$-values to be {1000, 800, 400, 50} in this work.

TABLE IV
AVERAGE PSNR, SSIM AND HFEN VALUES OF RECONSTRUCTING 31 TEST IMAGES AT RADIAL SAMPLING TRAJECTORIE

| $\alpha$-values | $R = 10$, Radial |
|---|---|
| 1200, 1000, 800, 400 | 29.83/0.8074/1.51 |
| 1000, 800, 400, 50 | **30.35/0.8265/1.40** |
| 800, 400, 50, 10 | 29.43/0.7939/1.64 |

To evaluate the influence of the noise level, experiments were performed with four different noise levels on MRI reconstruction. The quantitative results are summarized in Table V. As can be observed, setting noise level too high or too low is not suitable. Thus, $\sigma_\eta = 25$ is chosen in this work.

TABLE V
AVERAGE PSNR AND SSIM VALUES OF RECONSTRUCTING 31 TEST IMAGES AT Cartesian SAMPLING TRAJECTORIE

| $\sigma_\eta$ | 15 | 20 | 25 | 30 |
|---|---|---|---|---|
| $R$=6.7, 1D Cartesian | 28.51/0.7948 | 28.33/0.7853 | **28.88/0.8011** | 28.02/0.7731 |

### C. Convergence Analysis

Fig. 13 depicts the performance evolution with regard to iteration numbers in HF-DAEP. It can be seen that PSNR values converge to a stable position with an increase of iteration numbers. This observation demonstrates that the proposed method can efficiently optimize the energy functions to a satisfactory solution.

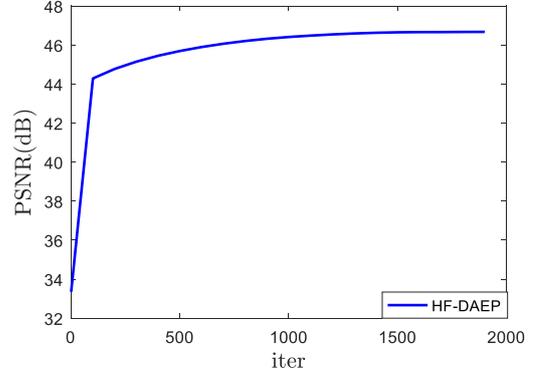

Fig. 13. Convergence tendency of HF-DAEP in CT imaging.

## VI. CONCLUSIONS

By analyzing the whole procedure in DL and CSC, we found that image decomposition and component summation can be seen as an approximation of forward and inverse transform. We proposed a multi-profile high-frequency transform-guided DAEP with a new strategy mainly acting on high-frequency components in images. Both qualitative and quantitative experimental results on highly under-sampled MRI and sparse-view CT reconstruction verified its superior performance.

The transformation is not limited to apply in DAEP, it is can be employed to other tools for prior information representation in forthcoming study. Furthermore, extending the proposed prior to other imaging modalities is also an interesting direction.